\documentclass[12pt]{iopart}
\usepackage{amssymb}
\usepackage{cite}
\usepackage{iopams}
\usepackage{bm}
\usepackage{upgreek}
\usepackage{graphicx}

\newcommand{\be}{\begin{equation}}
\newcommand{\ee}{\end{equation}}

\begin{document}

\title{Casimir friction between  polarizable particle and half-space with radiation and image damping at zero temperature}

\author{J. S. H{\o}ye$^{1}$, I. Brevik$^{2}$ and K. A. Milton$^3$}

\address{$^1$Department of Physics, Norwegian University of Science and Technology, 7491 Trondheim,  Norway}
\address{$^2$Department of Energy and Process Engineering, Norwegian University of Science and Technology, 7491 Trondheim, Norway}
\address{$^3$Homer L. Dodge Department of Physics and Astronomy,
University of Oklahoma, Norman, OK 73019 USA}
\ead{iver.h.brevik@ntnu.no}


\begin{abstract}

Casimir friction between a polarizable particle and a semi-infinite space is a
delicate physical phenomenon, as it concerns the interaction between a
microscopic quantum particle and a semi-infinite reservoir. Not unexpectedly,
results obtained in the  past about the friction force obtained via different
routes  are sometimes,  at least apparently,  wildly different from each other.
 Recently, we considered the Casimir friction force for two dielectric
semi-infinite plates moving parallel to each other [J. S. H{\o}ye and
I. Brevik, Eur. Phys. J. D {\bf 68}, 61 (2014)], and managed  to get essential
 agreement with results obtained by  Pendry (1997), Volokitin and Persson
(2007),  and Barton (2011). Our method was based upon use of the Kubo
formalism.  In the present paper we focus on the interaction between a
polarizable particle  and a dielectric half-space again, and calculate the
friction force using the same basic method as before.  The new ingredient in
the present analysis is that we take into account radiative  damping, and
derive the modifications thereof.  Some comparisons are also made with  works
from others. Essential agreement with the results of Intravaia, Behunin, and Dalvit
can also be achieved using the modification of the atomic polarizability by the
metallic plate.

\end{abstract}

\today

\maketitle

\section{Introduction} \label{sec1}

The Casimir friction problem - discussed extensively in recent years - is a
delicate problem in fundamental physics.  The typical geometrical setup is that of
two parallel  dielectric half-spaces  moving longitudinally with respect to each
other with constant velocity at  close spacing. In the limit when the material
density of one of the plates goes to zero, one obtains effectively the situation
where one single particle moves parallel to a dielectric half-space.

There are essentially two different avenues that one can follow to deal with
the problem. One, probably  the most conventional one,  is to make use of
field-theoretical methods. A second way is to make use of statistical
mechanical methods from the beginning. This is the method that two of us
 have been using ourselves in several papers; it implies use of the Kubo
response formula. In our opinion the statistical mechanical method is quite
compact and useful in the Casimir context in general. In the present paper our
main purpose is to extend our recent study for dielectric half-spaces
\cite{hoye14,hoye14a} to the case where the influence from
{\it radiative damping} is taken into account. However, the presence of the dielectric half-space is an additional source of energy dissipation that turns out to outweigh the radiative damping in free space by far at close separation. As mentioned below, this will be considered in two appendices.
We limit ourselves to the case of a moving polarizable particle close to a
half-space, and assume zero temperature, $T=0$.   Use of the effective
polarizability for a harmonic oscillator will now be needed. The oscillator is
damped by the radiation reaction due to the energy loss from radiated
electromagnetic dipole radiation. Whereas the solution of this classical problem is
 well known, we mention that  some time ago it was rederived by H{\o}ye and Stell
within the framework of  statistical mechanics for dielectric fluids \cite{hoye82}.
 It ought to be emphasized that  in the present work the polarizable particle with
its radiation damping will replace the influence from  one of the metal slabs with
its {\it finite conductivity} considered in Refs.~\cite{hoye14,hoye14a}. We work
out the consequences from use of a complex polarizability.

  As results found in the literature are sometimes very different, there is a need
for trying to make
contact between the various approaches. We mention that we have
earlier managed to get quite close agreement with  results obtained earlier by
Barton \cite{barton10A, barton10B,barton11,barton11B}, Pendry \cite{pendry97,
pendry98,pendry10}, and Volokitin and Persson \cite{volokitin99,volokitin03,
volokitin07,volokitin08,volokitin11,volokitin14}. To our knowledge this has
not been made before. As for the specific problem considered in the present
paper, we have made contact with the recent calculation
of Intravaia {\it et al.} \cite{intravaia14}.  The damping mechanism there is
quite different, being due to the modification of the atomic polarizability by
the conducting plate, and in fact in the two Appendices of this paper, we rederive
their result based on this alternative mechanism.

It may be of interest to give a  few remarks about the historical development of
this field. It is probably fair to say that the research on Casimir friction  was
started by the 1978 paper of Teodorovich \cite{teodorovich78}. Our own research
on Casimir friction was initiated early, by considering the simple but basic system
 consisting of a pair of harmonic oscillators that interacted weakly and moved
slowly relative to each other. Both  static \cite{hoye92} and time-dependent
interactions \cite{hoye93} were considered. The first papers of Pendry from the
1990's \cite{pendry97,pendry98} were influential, and established a kind of
approach later followed by Volokitin and Persson, and others.  A number of  papers
from the last 5 years  are listed in Refs.~\cite{dedkov08,dedkov10,
dedkov11,dedkov12,philbin09,silveirinha13A,maghrebi13,pieplow13,kyasov14,henkel14,
hoye10,hoye11a,hoye11b,hoye12a,hoye12b,hoye13,intravaia15}. The research literature
is quite diversified; some papers have focused on the complicated situation where
the relative velocity is relativistic and even necessitating the inclusion of
Cherenkov-like  effects. We will, however, limit ourselves to the case where the
velocity is nonrelativistic. Even this apparently simple case turns out to be
intricate enough.

The next section surveys the kind of statistical mechanical formalism that we have
been developing in a series of papers. In section 3 we focus on the extension of
the theory implying use of the radiation damping term (instead of finite
conductivity for a plate) in the polarizability for the particle, pointing out how
the conventional proportionality of $v^3$ in the zero temperature friction force
becomes replaced by a proportionality of $v^5$ ($v$ is velocity). Some numerical
evaluations and comparisons to other works are given in section 4.
In Appendix A, we set up a field theoretic formulation, similar to that
of Intravaia {\it et al.} \cite{intravaia14}, and use it to derive not only
the radiation reaction result found here, but also the induced damping
friction found in Ref.~\cite{intravaia14} (apart from a numerical factor).  In Appendix B we rederive that same
result using the statistical mechanical method explained in the body of this paper.
Thus, there is essential agreement concerning the formul{\ae} for Casimir friction.

\section{Evaluation of friction}
\label{sec2}

For readability we shall first outline the main properties of the
statistical-mechanical Kubo formalism that we are pursuing. The geometric setup is
the conventional one:
the lower plate 1 with surface situated at $z=0$ is at rest, while the upper plate
2 with surface situated at $z=d$ moves with  nonrelativistic velocity $v_x=v$ in
the $x$ direction. We let the origin be located at the surface of the lower plate,
and let ${\bf r}_0$ denote the relative initial position between two particles  in
the plates. Thus ${\bf r}_0=\{x_1,y_1,z_1-z_2\}$, with $x_2=y_2=0$.
The friction force between a polarizable particle and a half-plane, as mentioned
above, can be obtained when one half plane (chosen to be the upper one) is taken in
 the dilute limit.

Let $\rho_1$ and $\rho_2$ be the particle densities in the two half-planes, and let
 $\hat\psi(z_0,{\bf k}_\perp)$ be the Fourier transform in the $xy$-plane of the
$\psi({\bf r})$ (${\bf r}={\bf r}(t)$) part of the interaction $\psi({\bf r}) s_1 s_2$ between a pair of oscillators. The ${\bf k}_\perp$ is the Fourier variable,
$z_0=z_1-z_2$, and $s_1$ and $s_2$ are the vibrational coordinates. The dissipated
energy $\Delta E$ can be written in the form (cf. Eq.~(23) in Ref.~\cite{hoye14})
\begin{equation}
\Delta E=\frac{\rho_1 \rho_2}{(2\pi)^2}\int\limits_{z_1>d, z_2<0}\hat\psi(z_0, {\bf
 k}_\perp)\hat\psi(z_0, -{\bf k}_\perp) J(\omega_v)\,d{\bf k}_\perp dz_1 dz_2.
\label{1a}
\end{equation}
As it can be shown that with electrostatic dipole interaction
\begin{equation}
\hat\psi(z_0, {\bf k}_\perp)\hat\psi(z_0, -{\bf k}_\perp)
=(2q^2)^2\left(\frac{2\pi e^{-q\vert z_0\vert}}{q}\right)^2,
\label{1b}
\end{equation}
where $q=\vert {\bf k}_\perp\vert$ and $\omega_v=k_xv$, $k_x$ being the component
of the Fourier variable in the $x$ direction,  the $z_1$ and $z_2$ integrations of
Eq.~(\ref{1a}) can  be performed.  With this the dissipated energy becomes
\begin{equation}
\Delta E={\rho_1 \rho_2}\int e^{-2qd} J(\omega_v)\,d{\bf k}_\perp.
\label{1c}
\end{equation}

Then consider the quantity $J(\omega_v)$ in Eq.~(\ref{1a}). The evaluation of it is
 more demanding, and involves use of the quantum mechanical response function for
a pair of oscillators. Moreover, the quantity $J(\omega_v)$ contains information
about  the specific model that we use for the motion. In order to  avoid other
contributions to the change in energy, apart from dissipation, the motion should be
performed in a {\it closed loop}.  Without loss in generality we can assume that
 the  motion takes place in the $x$ direction. Initially, at $t<-\tau-\alpha \tau$
where  $\tau \gg 1$, $\alpha \gg1$, we assume that the upper plate is at rest. In
the period $t \in (-\tau-\alpha \tau, -\tau)$, it moves to the left with a low
constant velocity $-v/\alpha$ ($v$ is constant and positive). Then, for
$t\in (-\tau, +\tau)$, it moves to the right with finite velocity $v$, whereafter
it moves to the left again  with low velocity $-v/\alpha$ until it returns to its
initial position at   $t=\tau +\alpha \tau$. In this way the closed loop motion of
the upper plate is brought about.

In the limit $\alpha\rightarrow\infty$ dissipation from the initial and final very
low velocities can be neglected, and it turns out that we can write $J(\omega_v)$
 as
\begin{equation}
J(\omega_v)=C_- I(\omega_-)+C_+ I(\omega_+),
\label{1e}
\end{equation}
where the coefficients $C_+$ and $C_-$ are in general, at finite temperature,
\begin{eqnarray}
\nonumber
C_\pm&=&\frac{H}{\hbar}\sinh\left(\frac{1}{2}\beta\hbar\omega_\pm\right)
\label{19}
\end{eqnarray}
with $\omega_\pm=|\omega_1\pm\omega_2|$  and
\begin{equation}
H=\frac{\hbar^2\omega_1\omega_2\alpha_1\alpha_2}{4\sinh(\frac{1}{2}\beta\hbar
\omega_1)\sinh(\frac{1}{2}\beta\hbar\omega_2)}.
\label{20}
\end{equation}
Here $\alpha_1$ and $\alpha_2$ are the polarizabilities of a pair of particles, one
 in each half-plane,   $\omega_1$ and $\omega_2$ are the corresponding
eigenfrequencies, and $\beta= 1/(k_BT)$.

As for the quantities $I(\omega)$ in Eq.~(\ref{1e}), it turns out that they take
the form
\begin{equation}
I(\omega)=\pi \tau \frac{\omega_v^2}{\omega}[\delta(\omega-\omega_v)+\delta(\omega+\omega_v)]
\end{equation}
for large $\tau\rightarrow\infty $.
At $T=0$ as assumed in this paper, $C_-=0$ by which the $J(\omega_v)$ for a pair of
 particles will be
\begin{equation}
J(\omega_v)=C_+ \pi\tau\frac{\omega_v^2}{\omega_+}[\delta(\omega_+-\omega_v)+\delta
(\omega_++\omega_v)],
\label{1}
\end{equation}
\begin{equation}
C_+=\frac{1}{2}\hbar \omega_1\omega_2\alpha_1\alpha_2.
\end{equation}

Now consider  higher densities of particles. Then,  the polarizabilities are
generally to be replaced by
\begin{equation}
2\pi\rho\balpha\rightarrow\frac{\varepsilon-1}{\varepsilon+1},
\label{4}
\end{equation}
where $\varepsilon$ is the permittivity and $\rho$ as before is the number density. As usual, we consider only local permittivity. The case of nonlocal permittivity (spatial dispersion) has also been studied and can sometimes be of significance; cf., for instance, Ref.~\cite{sernelius06}. 

In general the oscillators have a frequency distribution. Then we have the
replacement
\begin{equation}
\alpha_a\rightarrow\alpha_{Ia}(m_a^2)\,d(m_a^2), \quad m_a=\hbar\omega_a, \quad (a=
1,2).
\label{5}
\end{equation}
With integration elements $d(m_1^2)\,d(m_2^2)=4\hbar^2 m_1 m_2\,d\omega_1\,
d\omega_2$ this gives the following expression for $J(\omega_v)$  $(\omega_+=
|\omega_v|$)
\begin{equation}
J(\omega_v)=2\pi\tau|\omega_v|\hbar^3\int\limits_0^{\vert\omega_v\vert}\omega_1
\omega_2 m_1 m_2\alpha_{I1}(m_1^2)\alpha_{I2}(m_2^2)\,d\omega_1.
\label{7}
\end{equation}

The frequency spectrum follows from the imaginary part of $\alpha$ or expression
(\ref{4}) for higher densities. It is obtained via Eqs.~(27)--(29) of
Ref.~\cite{hoye13}. One can write
\begin{equation}
\alpha(K)=f(K^2), \quad K=i\hbar\omega.
\label{8}
\end{equation}
Then it can be shown that the function $f(K^2)$ satisfies the relation
\cite{hoye82a}
\begin{equation}
f(K^2)=\int\frac{\alpha_I(m^2)m^2}{K^2+m^2}\,d(m^2)
\label{9}
\end{equation}
with
\begin{equation}
\alpha_I(m^2)m^2=-\frac{1}{\pi}\Im f(-m^2+i\gamma),\quad m=\hbar\omega=-iK,\quad
\gamma\rightarrow 0+.
\label{10}
\end{equation}

With the Drude model for a metal the permittivity is given by
\begin{equation}
\varepsilon=1+\frac{\omega_p^2}{\xi(\xi+\nu)}
\label{11}
\end{equation}
where $\xi=-i\omega$, the $\omega_p$ is the plasma frequency, and $\nu$ represents
damping of plasma oscillations due to finite conductivity of the medium. The
frequency spectrum that follows from the imaginary part of Eq.~(\ref{4}) is via
Eqs.~(\ref{8})--(\ref{11}) for small $m$ ($\rightarrow0$) given by
\begin{equation}
\alpha_I(m_2^2) m_2^2=D m_2,\quad
D=\frac{\hbar\nu}{\rho(\pi\hbar\omega_p)^2}.
\label{12}
\end{equation}

\section{Radiative damping included}
\label{sec3}
After having presented this brief overview of our kind of statistical mechanical
theory  we now turn to the new element in the present paper, which is to include
radiative damping for the polarizable particle moving above the half-space. (For a
single particle finite conductivity and corresponding damping is obviously absent.)
 The damping results from outgoing radiation from the particle. The resulting
polarizability can be found, for instance, as Eq.~(17) in  the paper of H{\o}ye and
 Stell on dielectric fluids \cite{hoye82},
\begin{equation}
\alpha_e=\frac{\alpha}{1+\frac{2}{3}i
\left(\frac{\omega}{c}\right)^3\alpha}
\label{13}
\end{equation}
 (note that Gaussian units are used). The $\omega$ in the reference is the
$-\omega/c$ above where $c$ is the velocity of light. (The minus sign is related to
 the different choice of sign by underlying Fourier transform.) It should be noted that this damping is for a polarizable particle in free space. However, it will be modified due to the presence of a dielectric medium where energy can dissipate. This extension is considered in two appendices where different methods are applied. For close separation the influence from the dielectric medium turns out to dominate by far. In  Ref.~\cite{hoye82} expression (\ref{13}) was given a statistical mechanical
derivation, but it is consistent with the one obtained via the radiated power from
an oscillating dipole. Here $\alpha$ is the polarizability of the undamped
oscillator
\begin{equation}
\alpha=\frac{\alpha_0\omega_0^2}{\omega_0^2-\omega^2}.
\label{14}
\end{equation}
 For small frequencies  ($\omega\ll\omega_0$) needed here ($v\ll c$) the $\alpha$
can be replaced by $\alpha_0$ by which Eq.~(\ref{13}) becomes
\begin{equation}
\alpha_e\rightarrow\alpha_1=\frac{\alpha}{1+\frac{2}{3}i
\left(\frac{\omega}{c}\right)^3\alpha_0}=\alpha_0\left(1-\frac{2}{3}i m^3
\frac{\alpha_0}{(\hbar c)^3}+\cdots\right)=f(-m^2+i\gamma)
\label{14a}
\end{equation}
With Eqs.~(\ref{8}) and(\ref{10}) this gives the frequency distribution
($\gamma\rightarrow 0+$)
\begin{equation}
\alpha_{I1}(m_1^2)m_1^2=-\frac{1}{\pi}\Im f(-m_1^2+i\gamma)=Bm_1^3, \quad B=
\frac{2\alpha_0^2}{3\pi(\hbar c)^3}.
\label{15}
\end{equation}

Expressions (\ref{12}) and (\ref{15}) for the frequency distributions are to be
inserted in integral (\ref{7}). Inserting Eqs.~(\ref{12}) and (\ref{15}) in
Eq.~(\ref{7}) we obtain ($\omega_2=|\omega_v|-\omega_1$)
\begin{equation}
J(\omega_v)=
2\pi\tau|\omega_v|\hbar^5 BD\int\limits_0^{|\omega_v|}\omega_1^3\omega_2\,
d\omega_1=2\tau \omega_v^6 H_{PII}, \quad H_{PII}=\frac{\pi}{20}\hbar^5 BD.
\label{16}
\end{equation}
For clarity, an extra subscript II is given to quantities differing from
corresponding quantities calculated in Ref.~\cite{hoye14}.  The most notable
difference from Eq.~(52) in \cite{hoye14} is that Eq.~(\ref{16}) contains the
factor $\omega_v^6$ instead of $\omega_v^4 $ (the factor $\tau$ present in  $H_P$
in  \cite{hoye14} is a misprint).
Note that in Ref.~\cite{hoye14} two equal metal half-planes were considered by
which Eq.~(\ref{12}) was used for both while in the present case Eq.~(\ref{15}) is
used for one of them.
By further integration of $\omega_v^6=(k_x v)^6$ ($v=v_x$) over orientations one
has $\int k_x^6\,d\phi=k_\perp^6\int_0^{2\pi}\cos^6\phi\,d\phi=2\pi q^6(5/16)$
($q=k_\perp$) such that $k_x^6$ can be replaced by $5 q^6/16$, i.e. we can write
\begin{equation}
J(\omega_v)=2\tau v^6 H_{PII}\frac{5}{16}q^6,
\label{17}
\end{equation}
which replaces Eq.~(53) in \cite{hoye14}. The most characteristic difference is
that $v^6q^6$ replaces $v^4q^4$.

We can now calculate the dissipated energy per unit area from Eq.~(\ref{1c}) to be
\begin{equation}
\Delta E_{PII}=2\tau v^6 H_{PII}G_{PII},
\label{20a}
\end{equation}
where $H_{PII}$ is given by Eq.~(\ref{16}) above, and
\begin{equation}
G_{PII}={\rho_1\rho_2}\int\limits_0^\infty\frac{5}{16}q^6 \,e^{-2qd}\,2\pi q\,
dq=\rho_1\rho_2\frac{5\cdot315\pi}{2^7d^8}.
\label{19a}
\end{equation}
The friction force per unit area  becomes accordingly
\begin{equation}
F_{PII}=-\frac{\Delta E_{PII}}{2\tau v}=-H_{PII}G_{PII}v^5=-\rho_1\rho_2
\frac{315\pi^2\hbar^5}{2^9d^8}BDv^5.
\label{21}
\end{equation}

The results above were for two half-spaces.
Consistent with the result (\ref{21}) the force between a {\it particle}  moving at
a distance $z_0$ away from a half-space will be given by (extra subscript II
omitted from now on)
\begin{equation}
F=-\frac{A}{z_0^9}
\label{22}
\end{equation}
where the coefficient $A$ follows from Eq.~(\ref{21}) as
\begin{equation}
A=8\rho_2\frac{315\pi^2\hbar^5}{2^9}BDv^5=\frac{105\hbar\nu\alpha_0^2}{2^5\pi
\omega_p^2c^3}v^5.
\label{24}
\end{equation}
(We have here made use of the fact that the force for two half-spaces results from
integrating the single-particle-half-space force over all $z_0$, from $d$ to
infinity.) Recall that $B$ and $D$ are given by Eqs.~(\ref{12}) and (\ref{15})
respectively. Note also that these expressions are given in Gaussian units,
implying that the polarizability  $\alpha_0$ has the dimension cm$^3$.

We emphasize the difference between the present result (\ref{21}), $\propto v^5$,
and the earlier result $\propto v^3$ obtained in \cite{hoye14}, Eq.~(56),  for two
metal half-planes with use of the same dielectric function (\ref{11}). The reason
is the  polarizability (\ref{13}) for the low density medium. With its
$\omega^3$-behavior in the denominator this differs in a qualitative way from
expression (\ref{11}) that gives a corresponding $\omega$-behavior in expression
(\ref{4}). This has the consequence that the frequency distribution has the
$B m_1^3$ behavior of Eq.~(\ref{15}) while the latter has the $Dm_2$ one of
Eq.~(\ref{12}) ($m=\hbar\omega$). This difference in behavior reflects the rapid
vanishing of the radiation damping as the frequency decreases.

\section{Remarks on numerical magnitudes}

Various approaches to the Casimir friction problem sometimes  look very different
from each other. This may not be unexpected. Due to this it may be difficult to
check whether disagreements between results obtained are real or merely reflect
differences in formalism. Little effort has to our knowledge been made by others to
 clear these things up. As mentioned, our method is based upon the Kubo formalism
\cite{kubo59}.

To begin with, let us present some encouraging results for the case of two
half-spaces. Volokitin and Persson, in their 2007 review article
\cite{volokitin07}, found the $T=0$ friction force here called $F_{\rm VP}$, to be
$F_{\rm VP}=F/2$, where $F$ is our result \cite{hoye14,hoye13}. The expression
$F_{\rm VP}$ was  based upon a calculation of Pendry \cite{pendry97}, who found
$F_{\rm Pendry}=F_{\rm VP}/6$ (this slight mismatch is seemingly due to a trivial
calculational error in Pendry's paper). The same situation was considered also by
Barton \cite{barton11B}, who found $F_{\rm Barton}=12F_{\rm Pendry}$ (when a factor
 $\zeta(5)=1.037$ is disregarded). So, we find actually agreement at $T=0$,
\begin{equation}
F=F_{\rm Barton} \quad \rm (two ~half~spaces).
\end{equation}

Then turn to the case which is more relevant to the present case, namely  one
particle moving outside a half-space. It would here be of interest to make contact
with the numerical results obtained recently by Intravaia {\it et al.}
\cite{intravaia14}, who considered a situation of the same kind. Now, it is known
that the largest polarizabilities for the elements are found for  alkali atoms. We
choose rubidium as an example (the same element as considered in
Ref.~\cite{intravaia14}). The static ground-state polarizability
$\alpha_0 \,\,(5 ^2S_{1/2})$ is in this case given in Ref.~\cite{steck} as
\begin{equation}
\alpha_0=h\times 0.0794~\rm Hz/(V/cm)^2, \label{25}
\end{equation}
with $h=6.626\times 10^{-34}~$J\,s being Planck's constant. Thus, in ordinary SI
units,
\begin{equation}
\alpha_0=5.26\times 10^{-39}\, \rm F\,m^2, \label{26}
\end{equation}
where F=C/V means farad (the dimension given in Ref.~\cite{intravaia14} is
incorrect). Using the conversion formula between Gaussian polarizability
$\alpha_0\rightarrow\alpha_{0G}$ and SI polarizability $\alpha_0$
\begin{equation}
\alpha_{0G}=\frac{1}{4\pi\varepsilon_0}\alpha_0,
\label{27}
\end{equation}
with the permittivity of vacuum $\varepsilon_0=8.85\cdot 10^{-12}$\,F/m,
we get the Gaussian ground-state polarizability
\begin{equation}
\alpha_{0G}=47.3\times 10^{-24}\, \rm cm^3 =47.3 ~\rm \AA^3, \label{28}
\end{equation}
in terms of the convenient unit $1~{\rm \AA}=10^{-8}~$cm. (Most other elements
have a static polarizability of order $1~{\rm \AA}^3$ or less.)

As for the plasma frequency and the relaxation parameter we choose for definiteness
 the values for gold, $\hbar\omega_p=9.0~{\rm eV}$ and $\hbar\nu=35~{\rm  meV}$ or
 $\omega_p=1.36\times10^{16}~$rad/s and $\nu=5.32\times 10^{13}~$rad/s (cf., for
instance, Ref.~\cite{hoye03}). Then, choosing $v=340~$m/s we obtain from
Eq.~(\ref{24})
\begin{equation}
A= 1.19\times 10^{-122}~\rm{Nm}^9=-1.19\times 10^{-99}~\rm dyne\, cm^9. \label{29}
\end{equation}
Further choosing $z_0=10~$nm  
we obtain from Eq.~(\ref{22})  the force on
 a single particle to be
\begin{equation}
F=-1.19\times 10^{-50}\,\rm{N}=-1.19\times 10^{-45}\, \rm dynes. \label{30}
\end{equation}
This is far beyond detectability in practice, and it is far lower than the
force found in Ref.~\cite{intravaia14}. However, the latter result was found
for a silicon half-space instead of gold. For silicon the resistivity
$\rho=6.4\cdot10^2\,\Omega$m was used, and with the other numbers unchanged
the friction force $F\approx-1.3\cdot 10^{-20}$\,N was found.
Likewise for the same situation with use of the relation
$\rho=\nu/(\varepsilon_0\omega_p^2)$ our Eq.~(\ref{22}) together with
(\ref{24}) give the tiny friction force $F=-2.3\cdot10^{-40}$\,N.
Although larger than result (\ref{30}) it is still far from the result of
Ref.~\cite{intravaia14}.
However, as shown in the Appendix A,  our result  can also be derived
using the formalism of Ref.~\cite{intravaia14}, so we conclude that the
difference resides in the different atomic dissipation mechanism assumed,
and indeed, by using the modification of the polarization induced by the
metallic plate, we recover (essentially) the result of that reference as well.
This is verified by a calculation in Appendix B which again reproduces (essentially)
the result of Ref.~\cite{intravaia14} using the formalism given above.
Thus the controversy appears resolved; in practice, the mechanism based on the
induced particle dissipation by the interaction with the plate is by far larger
than that caused by radiation reaction, although below the level of detectability
with present techniques.

\bigskip
We may conclude our work as follows: Our intention has been to evaluate the friction force at $T=0$ for a polarizable particle moving parallel to a metal half-space at close separation, using the statistical mechanical Kubo formalism. First, friction connected with the radiation reaction in free space is found. Then, the influence from dissipation in the metal half-space is included in the appendices where also an independent method of derivation is used. At close separation the latter dissipation is found to outweigh the former by far. Moreover, the results obtained by our method are compared with results obtained by others, based upon independent and very different methods. Essential agreement is obtained.

\appendix
\section*{Appendix A: Field theory (source theory) approach}
\label{appA}
\setcounter{section}{1}
In this appendix we wish to make contact with the formalism employed
by Intravaia et al. \cite{intravaia14} and rederive
the above result for Casimir
friction between a polarizable atom and a metal plate.  Because the approach
seems quite different, and the derivation in Ref.~\cite{intravaia14}
is quite terse, we describe it in detail.  (Here we set $\hbar=c=1$
and use, except where otherwise noted, rationalized Heaviside-Lorentz units.)

We start from the action term describing the interaction between two
polarization sources,
\begin{equation}
W=\frac12\int (\rmd x)(\rmd x')
\mathbf{P}(x)\cdot\bGamma(x,x')\cdot\mathbf{P}(x'),\label{action}
\end{equation}
where the integration is over all space-time coordinates, and $\bGamma$ is
the Green's dyadic for the system. The atom offers
a realization of the source $\mathbf{P}$;
if the atom moves with velocity $\mathbf{v}$,
as before, parallel to the surface, the change in the polarization in a
small time $\delta t$  is $\delta\mathbf{P}=-\delta t\mathbf{v}\cdot\bnabla
\mathbf{P}$.  The resulting variation in the action allows us to read off
the force due to the quantum friction in terms of the power ($T$ is the
``infinite'' time interval the system exists)
\begin{equation}
P=\mathbf{F\cdot v}=-\frac{\delta W}{T \delta t},
\ee
where
\begin{equation}
\fl\mathbf{F}=
\int \rmd(t-t')(\rmd\mathbf{r}_\perp)(\rmd\mathbf{r}_\perp')\rmd z\,\rmd z'
\mathbf{P}(\mathbf{r}_\perp,z,t)
\cdot(\bnabla)\bGamma(\mathbf{r}_\perp,z;\mathbf{r}'_\perp,z';
t-t')\cdot\mathbf{P}(\mathbf{r}_\perp',z',t').
\end{equation}
Now we do a $2+1$ dimensional break-up of the Green's dyadic for the surface,
\begin{equation}
\fl\bGamma(\mathbf{r}_\perp,z;\mathbf{r}'_\perp,z';t-t')=
\frac1{2\pi}\int_{-\infty}^\infty \rmd\omega\, \rme^{-\rmi\omega(t-t')}
\int\frac{(\rmd\mathbf{k}
_\perp)}{(2\pi)^2}\rme^{\rmi \mathbf{k}_\perp\cdot(\mathbf{r}_\perp
-\mathbf{r}_\perp')}\mathbf{g}(z,z';\mathbf{k}_\perp,\omega),
\end{equation}
and we replace the polarization by its realization by a moving dipole,
\begin{equation}
\mathbf{P}(\mathbf{r}_\perp,z,t)=\mathbf{d}(t)\delta(\mathbf{r_\perp-r_\perp}
(t))\delta(z-z_0),
\end{equation}
where the atom is a distance $z_0$ above the plate.  Here $\mathbf{r}_\perp(t)$ is
the atom's trajectory, which we approximate by motion with constant velocity,
$\mathbf{r}_\perp(t)=\mathbf{v}_\perp t$, in which case we obtain the simple
formula for the frictional force
\begin{equation}
\mathbf{F}=\int_{-\infty}^\infty \frac{\rmd\omega}{2\pi}
\int\frac{(\rmd \mathbf{k}_\perp)}
{(2\pi)^2}i\mathbf{k}_\perp\int^\infty_{-\infty} \rmd\tau\, e^{-\rmi(\omega-
\mathbf{k_\perp\cdot v_\perp})\tau}
\tr [\mathbf{d}(t')\mathbf{d}(t)\mathbf{g}(z_0,z_0)],
\end{equation}
with $\tau=t-t'$.  We interpret the product of dipole moments here as
a correlation function,
\begin{equation}
\mathbf{C}(\tau)=\frac12\langle \mathbf{d}(t)\mathbf{d}(t')+
\mathbf{d}(t')\mathbf{d}(t)\rangle,
\end{equation}
and we see the appearance of its Fourier transform, in the notation of
Ref.~\cite{intravaia14}
\begin{equation}
\mathbf{S}(\omega)=
\frac1{2\pi}\int_{-\infty}^\infty \rmd \tau\,\rme^{\rmi\omega\tau}
\mathbf{C}(\tau).
\end{equation}
Thus our final formula nearly
coincides with that given in Ref.~\cite{intravaia14}:
\begin{equation}
\mathbf{F}=\int_{-\infty}^\infty
\rmd\omega \int\frac{(\rmd\mathbf{k}_\perp)}{(2\pi)^2}
\rmi\mathbf{k}_\perp\tr [\mathbf{S}(\mathbf{\omega-k_\perp\cdot v_\perp})\cdot
\mathbf{g}(z_0,z_0;\mathbf{k}_\perp,\omega)].\label{fdtforce}
\end{equation}

The power spectrum $\mathbf{S}$
 is related to a generalized susceptibility through the
fluctuation-dissipation theorem (FDT).  We follow Landau and Lifshitz \cite{ll}
and write at zero temperature
\begin{equation}
\fl\mathbf{S}(\omega)=\mbox{sgn}(\omega)\frac1{2\pi}\Im \balpha(\omega),
\quad\mbox{and\,similarly}\quad\mathbf{g}(\omega)\to\mbox{sgn}(\omega)
\mathbf{g}(\omega).\label{s}
\end{equation}

The formula (\ref{fdtforce}) differs in two crucial aspects from that of
Ref.~\cite{intravaia14}: The integral ranges over both positive and negative
frequencies, and the sign of the argument of the power spectrum $\mathbf{S}$
is reversed.  These modifications are necessary to recover the correct results.
In fact, if we use (\ref{fdtforce}) to rederive the normal Casimir-Polder
force for
$\mathbf{v}=0$, by replacing $i\mathbf{k}_\perp\to \partial_z$, we see
immediately that the expected result is found,
\begin{equation}
F_z=-\partial_z U_{\rm CP},\quad U_{\rm CP}=-\int_{-\infty}^\infty \rmd\zeta
\tr\balpha_G(\rmi|\zeta|) \bGamma(z_0,z_0;\rmi|\zeta|),
\end{equation}
where we recognize that the relation between the Heaviside-Lorentz (used
elsewhere in this Appendix) and the
Gaussian values of the polarizability is $\balpha_{\rm HL}=4\pi \balpha_G$.
All that is needed to establish this identity is the Kramer-Kronig dispersion
relation
\begin{equation}
\alpha(i|\zeta|)=\frac2\pi\int_0^\infty \rmd\omega
\frac{\omega\Im\alpha(\omega)}{\omega^2+\zeta^2}.
\end{equation}

Then the frictional force can be shown to arise only from the low
frequency region because the $\mathbf{k}_\perp$ integrand must be even:
\begin{equation}
\mathbf{F}=\int\frac{(\rmd\mathbf{k}_\perp)}{(2\pi)^2}
\int_0^{\mathbf{k\cdot v}}\frac{\rmd\omega}{\pi}\mathbf{k}_\perp
\tr[\Im\balpha(\omega-\mathbf{k\cdot v})\Im \mathbf{g}(
\omega,k_\perp)].\label{f1}
\end{equation}
This result
differs from the formula given in Ref.~\cite{intravaia14} by a factor of
$\frac12$, and the sign of the argument of $\balpha$.

We now assume that the velocity of the atom is very small,
\begin{equation}
\omega\le\mathbf{k}_\perp\cdot\mathbf{v}_\perp\ll \omega_p,
\end{equation}
so that only the transverse magnetic (TM) part of the Green's dyadic contributes,
and for an isotropic atom we encounter only
\begin{equation}
\tr \Im\mathbf{g}(z_0,z_0;k,\omega)\approx k \Im\frac{\varepsilon-1}{\varepsilon+1}
\rme^{-2k z_0},\label{imgf}
\end{equation}
where
\begin{equation}
\Im\frac{\varepsilon(\omega)-1}{\varepsilon(\omega)+1}\approx \frac{2\omega
\nu}{\omega_p^2}.\label{imeps}
\end{equation}  This says that the dominant effects of Casimir friction
occur in the nonretarded evanescent regime.

All of this seems completely unambiguous, and noncontroversial.  It is the
nature of the susceptibility of the atom that that seems to cause some
consternation.  The assumption being made in the body of
this paper is that the dissipation
mechanism in the atom is that of radiation reaction, so as noted in the text
\begin{equation}
\Im\alpha(\omega)=\frac23\alpha_0^2\omega^3.\label{imalpha}
\end{equation}  Then when we substitute (\ref{s}), (\ref{imalpha}),
(\ref{imgf}), and (\ref{imeps}) into (\ref{fdtforce}), and carry out the
elementary integrations, we obtain (a factor of $4\pi$ comes from using
unrationalized Gaussian units)
\begin{equation}
F= -\frac{105}{32\pi}\frac{\alpha_0^2\nu}{\omega_p^2}\frac{v^5}{z_0^9}.\label{frr}
\end{equation}
This coincides with the result (\ref{24}) found in the text,
and indeed the integrals involved are the same as encountered there.

Quite different assumptions are made for the atomic susceptiblity in
Ref.~\cite{intravaia14}, leading to a very different dependence on atomic
velocity and distance, $F\propto v^3/d^{10}$.  In fact, it is easy to reproduce
that result.  The point is that the image of the charge distribution in the
atom lags behind that of the moving atom due to resistance in the metal.
This is captured by rewriting (\ref{action}) symbolically as
\begin{equation}
W=\frac12 \Tr \mathbf{E}\cdot \balpha_0\cdot\bGamma_0\cdot \balpha_0\cdot
\mathbf{E},\label{actionind}
\end{equation}
where $\balpha_0$ is the polarizability of the atom in empty space, and
$\bGamma_0$ is the Green's dyadic of the metal plate without the atom present.
Both are modified
in the presence of each other, and then the effective source  term is
\begin{equation}
W_{\rm eff}=\rmi\Tr \balpha\bGamma,
\quad \frac1{\rmi}\bGamma=\langle\mathbf{EE}\rangle.
\end{equation}
Therefore, the effective polarizability of the atom in the presence of the
plate is
\begin{equation}
\balpha_{\rm eff}=-\frac12\balpha_0\cdot\bGamma_0\cdot \balpha_0,
\end{equation}
or in the evanescent region, using (\ref{imeps}),
\begin{equation}
\Im\balpha_{\rm eff}
=-\frac1{\pi}\frac{\alpha_0^2}{(2z_0)^3}\frac{\omega\nu}{\omega_p^2}
\frac12\mbox{diag}(1,1,2),
\end{equation}
where
we have assumed that the atom is isotropic and possesses a real polarizability.
The anisotropy arises from that of the TM Green's dyadic.

Now we insert this into the general formula (\ref{fdtforce}) with (\ref{s}) for
the power
spectrum, together with the Green's dyadic properties on the principal diagonal
\begin{equation}
 \Im\mathbf{g}(z_0,z_0;k,\omega)\approx \frac{k}2
\Im\frac{\varepsilon-1}{\varepsilon+1}\rme^{-2k z_0}\frac12\mbox{diag}(1,1,2),
\label{imgf2}
\end{equation}
again with
(\ref{imeps}),
which immediately yields
\begin{equation}
F=-\frac{135\alpha_0^2v^3}{4\pi^3\sigma^2(2z_0)^{10}},\label{ibda}
\end{equation}
which is 3/8 times the result of Ref.~\cite{intravaia14}: This factor
comes from the trace over the product of Green's dyadics.
Here we use the connection between the conductivity of the metal and
the dissipation frequency in rationalized units,
\begin{equation}
\sigma=\frac{\omega_p^2}\nu.\label{cond}
\end{equation}.

Thus the controversy involving the friction between an atom and a metal surface
seems resolved.  The different results depend on the damping mechanism assumed.
 In practice, it would in fact appear that the last mechanism, found
by Ref.~\cite{intravaia14}, and confirmed, again  up to a factor, by
Ref.~\cite{intravaia15} is dominant for low velocities, although it still
seems far beyond experimental reach.

\section*{Appendix B: Statistical mechanics approach}
\label{appB}
\setcounter{section}{1}
We also want to make contact with the result of Ref.~\cite{intravaia14} using the
methods of Secs.~\ref{sec2} and \ref{sec3}. With radiation reaction the method of
Appendix A gave result (\ref{frr}) which fully agrees with results
(\ref{22}) and (\ref{24}).

The radiation reaction is an interaction on an oscillating dipole back on itself.
In Ref.~\cite{hoye82} it was found by statistical mechanical study of the
refractive  index of polarizable fluids that the radiating dipole interaction
gave a self-interaction that adds to the oscillator potential by which the given
polarizability is replaced by an effective polarizability.

In the present situation there is a dielectric half-plane near the polarizable
particle. This clearly absorbs radiation and thus adds to the damping of the
oscillator. To obtain this damping we need the field that the half-plane reflects
back to the  oscillator. For small separations it turns out that this field by far
outweighs the usual radiation reaction. Thus the latter can be neglected, and it
is sufficient to consider the electrostatic limit of the electric field on a
particle near near a  dielectric half-plane. As noted above Eq.~(\ref{actionind})
the solution of this problem (outside the half-plane) is equivalent to adding a
mirror oscillator in the half-plane. The electric field from a dipole moment
${\bf s}'$ is
\begin{equation}
{\bf E}=-\frac{s'}{r^3}(3({\bf \hat s}'\cdot{\bf \hat r}){\bf \hat r}-
{\bf \hat s}'),
\label{B1}
\end{equation}
where the carets denote unit vectors. Let ${\bf s}'$ be the mirror dipole of the
dipole moment ${\bf s}$ of the polarizable particle. For the components parallel
and transverse to ${\bf \hat r}$ we thus have (mirror picture has opposite
charges)
\begin{equation}
s_{||}'=\sigma s_{||}, \quad s_\perp'=-\sigma s_\perp,\quad \mbox{with} \quad
\sigma=\frac{\varepsilon-1}{\varepsilon+1}.
\label{B2}
\end{equation}
With this the electric field is equivalent to a self-interaction
\begin{equation}
\phi=-\int{\bf E}\cdot d{\bf s}=\frac{\sigma}{2r^3}(s_x^2+s_y^2+2s_z^2),
\label{B3)}
\end{equation}
with $s_z=s_{||}$, $s_\perp^2=s_x^2+s_y^2$, and where $r=2z_0$.

The oscillator with given polarizability $\alpha_0$ oscillates in a potential
\begin{equation}
\phi_0=\frac{1}{2\alpha_0}(s_x^2+s_y^2+s_z^2).
\label{B4}
\end{equation}
The resulting potential is thus $\phi_0+\phi$ with effective polarizability ($i=x,y,z$)
\begin{equation}
\frac{1}{\alpha_{ei}}=\frac{1}{\alpha_0}+\frac{\sigma}{(2z_0)^3}n_i,
\quad\mbox{with}\quad n_x=n_y=1,\quad n_z=2.
\label{B5}
\end{equation}
\begin{equation}
\alpha_{ei}=\alpha_0-\alpha_0^2\frac{\sigma}{(2z_0)^3}n_i+\cdots.
\label{B6}
\end{equation}
By the statistical  mechanical study only the situation with scalar polarizability was considered for simplicity. In the present case where the induced part is anisotropic, one will expect an effective scalar polarizability to be equivalent to some average of it. The approach of Appendix A with anisotropy included tells how this average should be performed. According to the Green's dyadic properties of Eq.~(\ref{imgf2}) double weight is put on the  $z$-component compared to the $x$- and $y$-components, i.e. the $z$-component contributes most to the friction as it does to the induced polarizability. Thus with weighted average $\langle n\rangle=(n_x+n_y+2n_z)/4=3/2$ the average of $\alpha_{ei}$ is
\begin{equation}
\alpha_{e}=\alpha_0-\alpha_0^2\frac{\sigma}{(2z_0)^3}\langle n\rangle+\cdots
\label{B7}
\end{equation}
The frequency spectrum is obtained like Eqs.~(\ref{14a}) and (\ref{15}) as
($m_1=\hbar\omega$)
\begin{equation}\fl
\alpha_{I1}(m_1^2)m_1^2=-\frac{1}{\pi}\left(-\frac{\alpha_0^2}{(2z_0)^3}\Im\sigma
\right)\langle n\rangle
=D_1m_1,\quad
D_1=\frac{3\alpha_0^2\hbar\nu}{8\pi z_0^3
(\hbar\omega_p)^2},
\label{B8}
\end{equation}
where Eq.~(\ref{imeps}) is used.
This frequency distribution is linear in $m_1$ like the one given by
Eq.~(\ref{12}) for the half-plane.
By evaluations similar to Eqs.~(\ref{16})--(\ref{24}) one
finds the friction force to be given by Eq.~(56) of Ref.~\cite{hoye14} with one
factor $\rho D$ replaced by $\rho_1 D_1$ where $\rho_1\rightarrow 0$ is the low
density of polarizable particles forming the other half-plane utilized in
Secs.~\ref{sec2} and \ref{sec3}
\begin{equation}
F_D=-\frac{15\pi^2}{64d^6}\rho_1 D_1 \rho D(\hbar v)^3
\label{B9}
\end{equation}
with $D$ given by Eq.~(\ref{12}). Consistent with this and similar to
Eqs.~(\ref{22})--(\ref{24}) the friction force on a single particle becomes
\begin{equation}
F=-\frac{A}{z_0^{7}} \quad \mbox{with} \quad A=6\cdot\frac{15\pi^2}{64}
D_1 \rho D(\hbar v)^3
=\frac{135\alpha_0^2(\hbar\nu)^2}{2^8\pi(\hbar\omega_p)^4 z_0^3}
(\hbar v)^3.
\label{B10}
\end{equation}
(By the
previous derivation the $F_D$ followed from $F$ with $D_1=D$ constant. Integration with varying $D_1$ would only change (\ref{B9}) by a factor 6/9 while (\ref{B10}) would remain unchanged.)
This coincides with the result (\ref{ibda}) with (\ref{cond}) and conversion
formula (\ref{27}) inserted (with units where $\varepsilon_0=1$ as used in Appendix A). Thus the result (up to the trace factor mentioned)
of Ref.~\cite{intravaia14}
 has been established in two rather independent ways where the arguments and
reasonings are somewhat different.

\ack
KAM thanks Diego Dalvit for helpful discussion on Casimir friction, and the
Julian Schwinger Foundation for financial support of this work.

\end{document}